\documentclass[11pt, onecolumn]{IEEEtran}

\usepackage{times,amsmath,epsfig,amsfonts,amssymb,graphicx,url,cite,subfigure,colortbl,bm,upgreek,soul}

\begin{document}
\title{Hybrid Beamforming for Massive MIMO -- \\ A Survey}
\author{Andreas F. Molisch~\IEEEmembership{Fellow,~IEEE,}
	Vishnu V. Ratnam~\IEEEmembership{Student Member,~IEEE,}
	Shengqian Han~\IEEEmembership{Member,~IEEE,}
	Zheda Li~\IEEEmembership{Student Member,~IEEE,}
	Sinh Le Hong Nguyen~\IEEEmembership{Member,~IEEE,}
	Linsheng Li~\IEEEmembership{Member,~IEEE}
        and~Katsuyuki Haneda~\IEEEmembership{Member,~IEEE}
\thanks{A. F. Molisch, V. V. Ratnam, S. Han and Z. Li are with the Ming Hsieh Department of Electrical Engineering, University of Southern California, Los Angeles, CA 90089-2565 USA (e-mail: molisch@usc.edu, ratnam@usc.edu, zhedali@usc.edu). S. Han is also with the School of Electronics and Information Engineering, Beihang University, Beijing, China (e-mail: sqhan@buaa.edu.cn).}
\thanks{S. L. H. Nguyen, L. Li and K. Haneda are with the Aalto University School of Electrical Engineering, Espoo, Finland (e-mail: sinh.nguyen@aalto.fi, katsuyuki.haneda@aalto.fi). L. Li is presently with Huawei Helsinki, Finland (e-mail: linsheng.li@huawei.com).}
\thanks{This paper with a comprehensive list of references is available at http://arxiv.org/abs/1609.05078.}}

\maketitle

\begin{abstract}
Hybrid multiple-antenna transceivers, which combine large-dimensional analog pre/postprocessing with lower-dimensional digital processing, are the most promising approach for reducing the hardware cost and training overhead in massive MIMO systems. This paper provides a comprehensive survey of the various incarnations of such structures that have been proposed in the literature. We provide a taxonomy in terms of the required channel state information (CSI), namely whether the processing adapts to the instantaneous or the average (second-order) CSI; while the former provides somewhat better signal-to-noise and interference ratio, the latter has much lower overhead for CSI acquisition. We furthermore distinguish hardware structures of different complexities. Finally, we point out the special design aspects for operation at millimeter-wave frequencies.

\end{abstract}


\section{Introduction}

Multiple-input multiple-output (MIMO) technology, i.e., the use of multiple antennas at transmitter (TX) and receiver (RX), has been recognized since the seminal works of Winters~\cite{Winters87_TC}, Foschini and Gans~\cite{Foschini98_WPC}, and Telatar~\cite{Telatar99_ETT}, as an essential approach to high spectral efficiency (SE). In its form of multi-user MIMO (MU-MIMO), it improves SE in two forms: (i) a base station (BS) can communicate simultaneously with multiple user equipments (UEs) on the same time-frequency resources, (ii) multiple data streams can be sent between the BS and each UE.
The total number of data streams (summed over all UEs in a cell) is upper limited by the smaller of the number of BS antenna elements, and the sum of the number of all UE antenna elements.

While MU-MIMO has been studied for more than a decade, the seminal work of Marzetta~\cite{Marzetta10_TWC} introduced the exciting new concept of ``massive MIMO", where the number of antenna elements at the BS reaches dozens or hundreds. Not only does this allow to increase the number of data streams in the cell to very large values, but it also simplifies signal processing, creates ``channel hardening" such that small-scale fading is essentially eliminated, and reduces the required transmission energy due to the large beamforming gain; see, e.g., \cite{Larsson14_CM} for a review. Massive MIMO is {\em beneficial} at {centimeter-wave} frequencies, but is {\em essential} in the millimeter-wave bands,\footnote{In a slight abuse of notation, we denote 1-10 GHz as ``cm-waves'', and 10-100 GHz as ``mm-waves"} since the high free-space pathloss at those frequencies necessitates large array gains to obtain sufficient signal-to-noise ratio (SNR), even at moderate distances of about $100$ m.

Yet the large number of antenna elements in massive MIMO also poses major challenges: (i) a large number of radio frequency (RF) chains (one for each antenna element) increases cost and energy consumption; (ii) determining the channel state information (CSI) between each transmit and receive antenna uses a considerable part of spectral resources.

A promising solution to these problems lies in the concept of {\em hybrid} transceivers, which use a combination of analog beamformers in the RF domain, together with digital beamforming in baseband, connected to the RF with a smaller number of up/downconversion chains. This concept was first introduced in the mid-2000s by one of the authors and collaborators in \cite{Zhang05_TSP, Sudarshan_et_al_2006}. It is motivated by the fact that the number of up-downconversion chains is only lower-limited by the number of data streams that are to be transmitted; in contrast, the beamforming gain and diversity order is given by the number of antenna elements if suitable RF beamforming is done. While formulated originally for MIMO with arbitrary number of antenna elements (i.e., covering both massive MIMO and small arrays), the approach is of interest in particular to massive MIMO. 
Interests in hybrid transceivers has therefore accelerated over the past three years (especially following the papers of Heath and co-workers, e.g., \cite{Alkhateeb14_CM}), where various structures have been proposed in different papers. The time thus seems ripe for a review of the state of the art, and a taxonomy of the various transceiver architectures (often simplified to provide computational or chip-architectural advantages) and algorithms. The current paper aims to provide this overview, and point out topics that are still open for future research.

This survey covers hybrid beamforming structures using instantaneous or average CSI in Sections~II and III, respectively. A special structure incorporating switches between the analog and digital parts is described in Sec. IV.
Sec.~V clarifies constraints at mm-wave bands due to propagation conditions and hardware imperfections.
A summary and conclusions in Sec. VI round out the paper.

\section{Hybrid Beamforming Based on Instantaneous CSI}
\begin{figure}[!htb]
\centering
\includegraphics[width=0.85\textwidth]{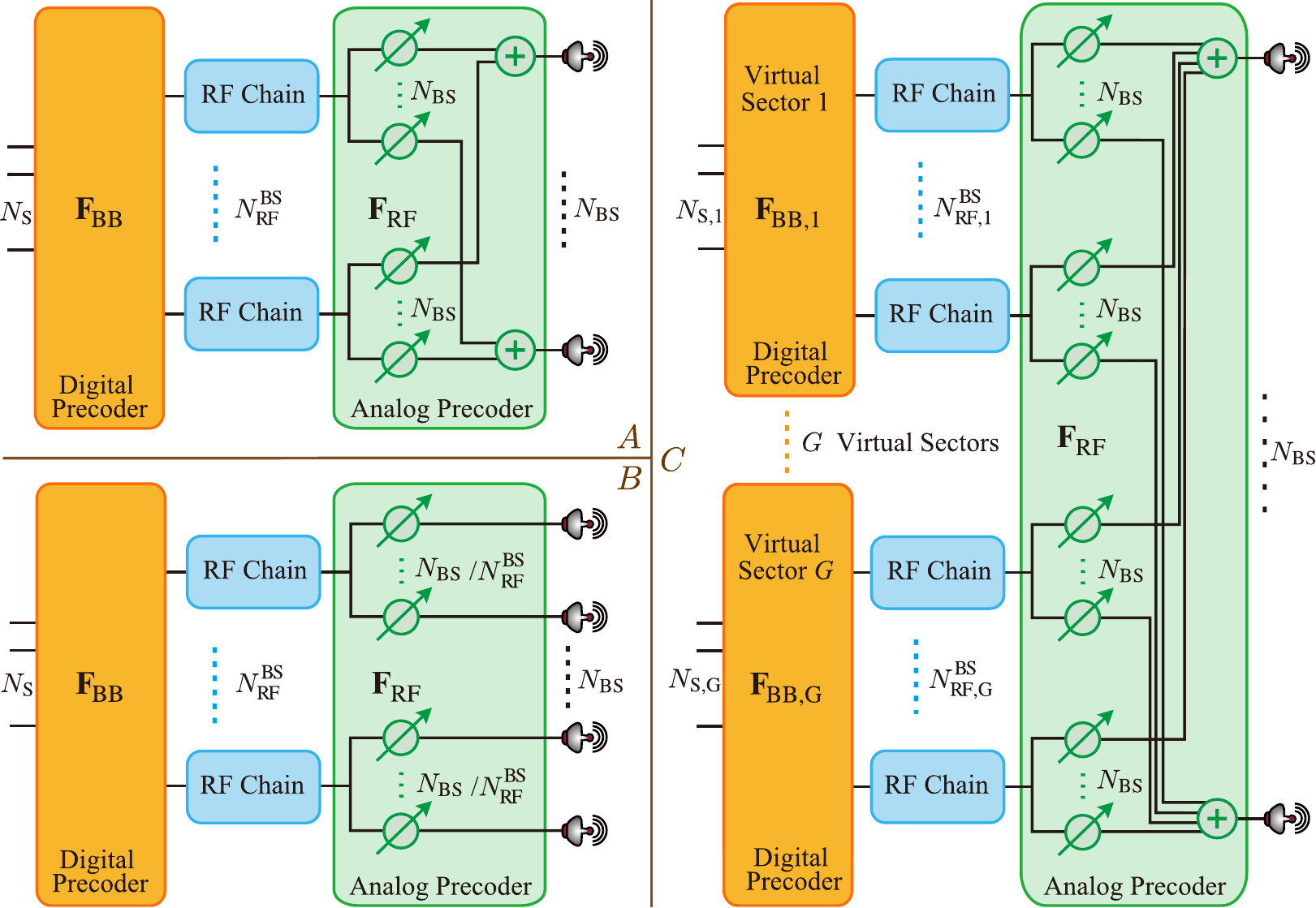}
\caption{{Block diagrams of hybrid beamforming structures at BS for a downlink transmission, where structures $A$, $B$, and $C$ denote the full-complexity, reduced-complexity, and virtual-sectorization structures, respectively.}}
\label{Fig:HybridStructure}
\end{figure}

{Fig. \ref{Fig:HybridStructure} shows block diagrams of three hybrid beamforming structures at the BS, where we assume a downlink transmission from the BS (acting as TX) to the UE (RX). {The classification is applicable to both cm- and mm-wave bands.} At the TX, a baseband digital precoder $\mathbf{F}_{\rm BB}$ processes $N_{\rm S}$ data streams to produce $N_{\rm RF}^{\rm BS}$ outputs, which are upconverted to RF and mapped via an analog precoder $\mathbf{F}_{\rm RF}$ to $N_{\rm BS}$ antenna elements for transmission. The structure at RX is similar: an analog beamformer $\mathbf{W}_{\rm RF}$ combines RF signals from $N_{\rm UE}$ antennas to create $N_{\rm RF}^{\rm UE}$ outputs, which are downconverted to baseband and further combined using a matrix $\mathbf{W}_{\rm BB}$, producing signal $\mathbf{y}$ for detection/decoding.\footnote{Obviously, a UE with a single antenna element is a special case.} Hence, we use terms ``beamformer" and ``precoder/combiner" interchangeably hereinafter. For a full-complexity structure, Fig. \ref{Fig:HybridStructure}$A$, each analog precoder output can be a linear combination of {\em all} RF signals. Complexity reduction at the price of a somewhat reduced performance can be achieved when each RF chain can be connected only to a subset of antenna elements~\cite{14-Xu2015}, as in Fig. \ref{Fig:HybridStructure}$B$. Different from structure $A$ and $B$ where baseband signals are jointly processed by a digital precoder, structure $C$ employs the analog beamformer to create multiple ``virtual sectors'', which enables separated baseband processing, downlink training, and uplink feedback and therefore reduces signaling overhead and computational complexity~\cite{05-Adhikary2013}.}

Even assuming full-instantaneous CSI at the TX, it is very difficult to find the analog and digital beam forming matrices that optimize, e.g., the net data rates of the UEs \cite{25-ElAyach2014}. The main difficulties include:
\begin{itemize}
\item
Analog and digital beamformers at each link end, as well as combiners at the different link ends, are coupled, which makes the objective function of the resulting optimization non-convex.
\item
Typically the analog precoder/combiner is realized as a phase-shifter network, which imposes additional constraints on the elements of  $\mathbf{W}_{\rm RF}$ and $\mathbf{F}_{\rm RF}$.
\item
Moreover, with finite-resolution phase shifters, the optimal analog beamformer lies in a discrete finite set, which typically leads to NP-hard integer programming problems.
\end{itemize}

Two main methodologies are explored to alleviate these challenges and achieve feasible near-optimal solutions.

\subsection{Approximating the optimal beamformer} \label{sec_2A}

For single-user MIMO (SU-MIMO), we start with optimum beamforming for the fully digital case with $N_{\rm RF}^{\rm BS}=N_{\rm BS}$ and $N_{\rm RF}^{\rm UE}=N_{\rm UE}$, where the solution is known (dominant left/right singular vectors of a channel matrix $\mathbf{H}$ from singular value decomposition (SVD)). Then, \cite{14-Xu2015, 25-ElAyach2014, 24-Ni2015} find an (approximate) optimum hybrid beamformer by minimizing the Euclidean distance to this fully digital one. The objective function of the approximation problem is still non-convex, but much less complex than the original one. For sparse channels (as occur in mm-wave channels), minimizing this distance provides a quasi-optimal solution~\cite{25-ElAyach2014}. In non-sparse channels, such as usually occur at {cm-wave} bands, an alternating optimization of analog and digital beamformer can be used. A closed-form solution for each of the alternating optimization steps can be developed (i) for the reduced-complexity structure~\cite{14-Xu2015}, while (ii) for the full-complexity structure~\cite{24-Ni2015}, the non-convex problem can be expanded into a series of convex sub-problems by restricting the phase increment of the analog beamformer within a small vicinity of its preceding iteration. Ref.~\cite{Zhang05_TSP} provides formulations (and approximate solutions) for both the diversity and the spatial multiplexing case. Refs.~\cite{Venkateswaran2010-ICASSP,Venkateswaran2010-TSP} explore the minimum mean square error (MMSE) hybrid receiver design and investigate the impact of analog-digital-converter (ADC) with different resolutions. 

  \begin{figure}[!htb]
    \centering
    \includegraphics[width=0.45\textwidth]{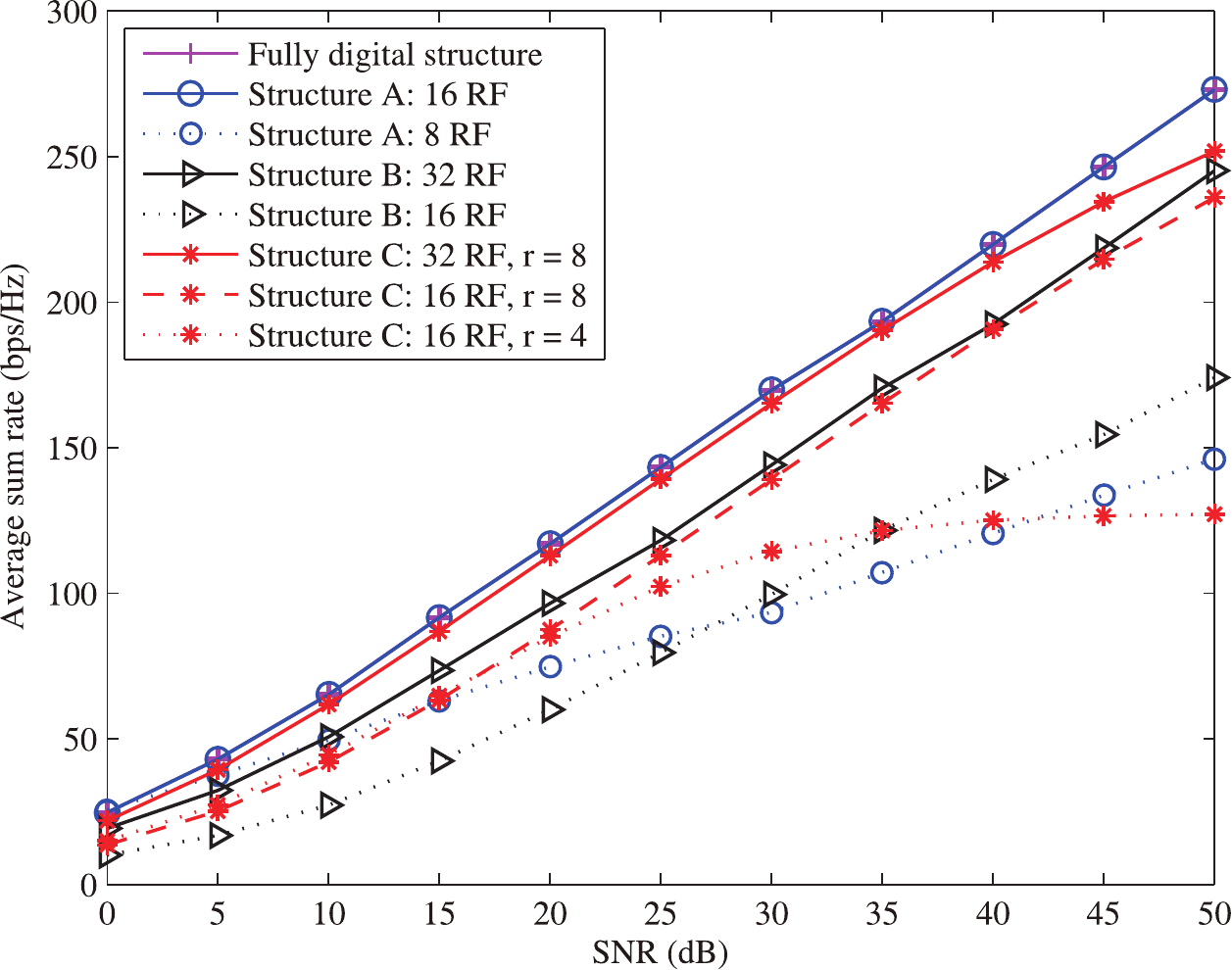}
    \caption{Performance comparison of the three hybrid structures with MU-MIMO; $N_{\rm BS}=64$, $N_{\rm UE}=1$, 4 groups of users located in a sector with mean directions $[-45^\circ, -15^\circ, 15^\circ, 45^\circ]$, and each group has $4$ users. AoDs of MPCs concentrate around the mean directions of each group with $10^\circ$ AoD spread. {This analysis assumes ideal hardware and typical channel conditions for cm-waves.}}
    \label{Fig:3structure}
  \end{figure}

Figure \ref{Fig:3structure} compares the performance of the three structures for downlink transmission of single-cell multi-user (MU) massive MIMO. The full-complexity structure of Fig.~\ref{Fig:HybridStructure}A performs the same as the fully digital structure when the number of RF chains is no smaller than the number of users (or streams). Performance loss of structure B is rather large for the considered MU case, though it is much smaller for SU-MIMO (not shown here). For structure C, the employed algorithm (JSDM, see also Sec. III) divides the users into 4 or 8 groups, which might lead to a performance floor due to inter-group interference; note that the significantly reduced training overhead of JSDM is not shown here; see Sec. III for discussion.   

\subsection{Decoupling the design of the analog and digital beamformers}

One of the main challenges in hybrid beamformer design is the coupling among analog and digital beamformers, and between the beamformers at TX and RX. This motivates decoupling the beamformer designs for reducing the problem complexity.
By assuming some transceiver algorithms, optimization of beamforming matrices can be solved sequentially. For example, in order to maximize the net rate for SU-MIMO, one can eliminate the impact of the combiner on the precoder by assuming a fully digital mean square error minimization (MMSE) receiver~\cite{20-Sohrabi2015}. Further decoupling of the analog and digital precoder is possible by assuming that the digital precoder is unitary. Subsequently, $\mathbf{F}_{\rm RF}$ is optimized column by column by imposing the phase-only constraint on each antenna. With the known analog precoder, a closed-form expression of the digital precoder can then be obtained.
Alternatively, some simple heuristic decoupling beamforming strategies have been explored. For example, the element-wise normalized conjugate beamformer can be used as the analog precoder~\cite{07-Liang2014, 04-Ying2015}, with which the asymptotic signal-to-interference-plus-noise ratio (SINR) of hybrid beamforming is only reduced by a factor of $\frac{\pi}{4}$ compared to fully digital beamforming when letting the number of antenna elements and streams, $N_{\rm BS}$ and $N_{\rm S}$, go to infinity while keeping $\frac{N_{\rm BS}}{N_{\rm S}}$ constant.

Extending to the situation where the UE is also equipped with a hybrid structure for MU-MIMO, one can first construct the RF combiner by selecting the strongest receive beams from the Fourier codebook to maximize the Frobenius norm of the combiner-projected channel~\cite{Ni15-arXiv2}. Then, the same normalized eigenbeamformer is implemented as the analog precoder on the effective channel~\cite{07-Liang2014, 04-Ying2015}. In the baseband, the BS performs block diagonalization (BD) over the projected channel to suppress inter-user interference.

\subsection{Wideband hybrid beamforming}

The above discussion focused on narrowband (i.e., single-subcarrier) systems. In wideband orthogonal frequency division multiplexing (OFDM) systems, however, analog beamformers cannot have different weights across subcarriers; for strongly frequency selective channels, such beamformers extending over the whole available band adapt to the average channel state (compare Sec.~III).

Frequency-domain scheduling such as the one in~\cite{27-Kim2013} was believed unnecessary for fully digital massive MIMO systems because the sufficiently large number of antennas can harden the channels and provide sufficient spatial degrees of freedom for multiplexing UEs~\cite{Larsson14_CM}. However, under practical constraints on array size (e.g., according to 3GPP LTE Release 13~\cite{36897}), frequency-domain scheduling is still necessary for hybrid transceivers~\cite{Kong2015, Bogale15-arXiv}. With frequency-domain scheduling, UEs are served on different subcarriers, making the existing narrowband hybrid precoders no longer applicable. Existing works~\cite{Kong2015, Geng13_GLOBECOM} have studied the joint optimization of wideband analog precoder and narrowband digital precoders, aimed at minimizing the BS transmit power or maximizing the sum rate of UEs.

Another important issue in the existing design of hybrid beamforming is control signaling coverage. While narrow analog beams are preferred for user-specific data transmission, wide beams are preferred for broadcasting control signals to all UEs. This problem may be solved, e.g., by splitting signaling and data planes so that they are transmitted at different carrier frequencies. For the massive MIMO systems operated at {cm-wave} frequencies, however, the splitting architecture is not necessarily needed, see, e.g.,~\cite{Kong2015-ICCC}.

\subsection{Impact of phase-only constraint and the number of RF chains}

Hybrid beamforming does not necessarily perform inferior to fully digital beamforming. {The analog beamforming can be implemented by means of phase shifters together with variable gain amplifiers. In this case, analog beamforming can provide the same functionality as digital beamforming, and combine desired multipath components (MPCs) (and suppress interfering MPCs) to the same degree as linear digital processing. Thus, in a narrowband massive MIMO system, with full-instantaneous CSI at the TX, this hybrid beamforming can achieve the same performance as fully digital beamforming, if $N_{\rm S} \le N_{\rm RF}$~\cite{Zhang05_TSP}.
A similar result can be obtained for a wideband system, where the number of RF chains of the hybrid structure should be not smaller than $\min(N_{\rm BS}, N_{\rm S,wb})$ with $N_{\rm S,wb}$ denoting the total number of data streaming over all subcarriers~\cite{Kong2015}.

Since two phase-only entries for the analog precoder are equivalent to a single unconstrained (amplitude and phase) entry, fully digital performance can be achieved with phase-only hybrid structures if $N_{\rm RF}^{\rm BS}\geq 2 N_{\rm S}$ in narrowband systems~\cite{Zhang05_TSP, 20-Sohrabi2015}}.

\section{Hybrid Beamforming Based on Averaged CSI}
\subsection{Average CSI based Hybrid Beamforming}
A major challenge for the beamformers discussed in Sec. II is the overhead for acquiring CSI at the BS. Information-theoretic results taking training overhead into account show that for time-division duplexing (TDD) systems, the spatial multiplexing gain (SMG) of massive MIMO downlinks with fully digital structure equals $M(1-\frac{M}{T})$, where $M=\min{(N_{\rm BS},K,\frac{T}{2})}$, $K=N_{\rm S}$ is the number of single-antenna users, and $T$ is the number of channel uses in a coherence time-frequency block \cite{05-Adhikary2013}. {In frequency-division-duplexing (FDD) systems, the overhead is even larger, since both downlink training and uplink feedback for each antenna are required.}  {In addition to the coherence time, the frame structure of systems may provide additional constraints for the pilot repetition frequency and thus the training overhead.}

It is evident that for any massive MIMO systems relying on full CSI between all antenna elements of the BS and UEs, the maximal achievable SMG is limited by the size of the coherence block of the channel because $N_{\rm BS}$ and $K$ are generally large. This necessitates the design of transmission strategies with reduced-dimensional CSI to relieve the signaling overhead. Specifically, recent research has considered analog beamforming based on slowly-varying second order statistics of the CSI at the BS {(a two-stage beamformer, with the first analog stage based on the average CSI only, followed by a digital one adapted to instantaneous CSI)}. The beamforming significantly reduces the dimension of the effective instantaneous CSI for digital beamforming within each coherent fading block by taking advantage of a small angular spread at the BS. Such structures work robustly even with the analog beamformers that cannot usually adapt to varying channels as quickly as digital beamformers. 

Hybrid beamformers using average CSI for the analog part were first suggested in \cite{Sudarshan_et_al_2006}, which also provided closed-form approximations for the optimum beamformer in SU-MIMO systems. For the MU case, \cite{05-Adhikary2013} proposed a scheme called ``Joint Spatial Division Multiplexing" (JSDM), which considered a hybrid-beamforming BS and single-antenna UEs; to further alleviate downlink training/uplink feedback burden, UEs with similar transmit channel covariance are grouped together and inter-group interference is suppressed by an analog precoder based on the BD method.

Specifically, using the Karhunen-Loeve representation, the $N_{\text{BS}}$-by-$1$ channel vector can be modeled as $\mathbf{h}=\mathbf{U}\mathbf{\Lambda}^{\frac{1}{2}}\mathbf{w}$,
where $\mathbf{w}\in\mathbb{C}^{r\times 1}\sim\mathcal{CN}(\mathbf{0},\mathbf{I}_r)$, $\mathbf{\Lambda}$ is an $r$-by-$r$ diagonal matrix, which aligns eigenvalues of channel covariance $\mathbf{R}$ on its diagonal, $\mathbf{U}\in\mathbb{C}^{M\times r}$ indicates the eigenmatrix of $\mathbf{R}$, and $r$ denotes the rank of the channel covariance. Dividing UEs into $G$ groups and assuming that UEs in the same group $g$ exhibit the same channel covariance $\mathbf{R}_g$ with rank $r_g$, the JSDM analog precoder~is
\begin{align}
&\mathbf{F}_{\text{RF}}=[\mathbf{F}_{\text{RF},1}, \dots, \mathbf{F}_{\text{RF},G}] \ \text{with}\ \mathbf{F}_{\text{RF},g}=\mathbf{E}_{g}\mathbf{G}_g.
\end{align}
By selecting $r_g^\star\leq r_g$ dominant eigenmodes of $\mathbf{R}_g$, denoted by $\mathbf{U}_g^\star$, we build the eigenmatrix of the dominant interference to the $g$-th group: $\mathbf{\Xi}_g=[\mathbf{U}_1^\star,...,\mathbf{U}_G^\star]$. Then, $\mathbf{E}_{g}$ consists of the null space of $\mathbf{\Xi}_g$, and $\mathbf{G}_g$ consists of dominant eigenvectors of $\mathbf{E}_{g}^{\dag}\mathbf{R}_g\mathbf{E}_{g}$.
This creates multiple ``virtual sectors" in which downlink training can be conducted in parallel, and each UE only needs to feed back the intra-group channels, leading to the reduction of both training and feedback overhead by a factor equal to the number of virtual sectors.

In practice, however, to maintain the orthogonality between virtual sectors, JSDM often conservatively groups UEs into only a few groups, because UEs' transmit channel covariances tend to be partially overlapped with each other. This limits the reduction of training and feedback overhead. Once grouping UEs into more virtual sectors violates the orthogonality condition, JSDM is not able to combat the inter-group interference. Eliminating overlapped beams of UEs in different groups is a heuristic approach to solve this problem~\cite{06-Adhikary2014}. 
In~\cite{Zheda2016}, JSDM is generalized to support non-orthogonal virtual sectorization, and a modified MMSE algorithm is proposed to optimize the multi-group digital precoders to maximize the lower bound of the average sum rate.

Two UE grouping methods have been proposed as extension to JSDM~\cite{07-Adhikary2014}:
K-means clustering~\cite{14-Xu2014}, or fixed quantization.
In the large antenna limit, the number of downlink streams served by JSDM can be optimized given the angle of departure (AoD) of MPCs and their spread for each UE group. To reduce the complexity of JSDM, in particular due to
SVD, an online iterative algorithm can be used to track the analog precoder under time-varying channels~\cite{32-Chen2014}.
When considering single-antenna UEs, Fourier codebook based analog precoder and zero-forcing (ZF) digital precoder, the performance of JSDM can be further improved by jointly optimizing the analog precoder and allocation of RF chains to groups based on second order channel statistics~\cite{09-Liu2014} . This principle can be extended to multicell systems, where an outage constraint on the UEs' SINR can be considered~\cite{10-Liu2015}.

\subsection{Decoupling of Analog and Digital Beamformers}

Different from Section II where both analog and digital beamformers are based on instantaneous CSI, now analog and digital beamformers are based on the average CSI and the instantaneous effective CSI, respectively. Thus, to find the optimal beamformers, one needs to first design the digital beamformer for each snapshot of the channel and then derive the analog beamformer based on their long-term time-average, making their mathematical treatment difficult. Decoupled designs of the analog and digital beamformers therefore make the optimization problem simpler and practically attractive. For SU-MIMO where a UE is equipped with a single RF chain and multiple antennas, the optimal analog combiner is intuitively the strongest eigenmode of the UE-side channel covariance. However, when there are more RF chains at the UE, the strongest eigenmodes are not always the optimal combiners since they may be associated with a single transmit eigenmode of the BS-side channel covariance. For MU-MIMO with multiple RF chains at both link ends, the digital beamformer design needs to consider the UE-level spatial multiplexing and inter-user interference suppression, which will affect the analog beamformer design. In~\cite{Zheda2016}, the optimality of decoupling analog and digital beamformers is shown under Kronecker channel model in the sense of maximizing the so-called ``intra-group signal to inter-group interference plus noise ratio''.

\subsection{Full-dimensional (FD) MIMO in 3GPP}

While the 3GPP standard does not prescribe particular transceiver architectures, HDA structures have motivated the design of CSI acquisition protocols in Release 13 of LTE-Advanced Pro in 3GPP~\cite{36897}, especially the non-precoded and beamformed pilots for FD MIMO. The non-precoded beamformer is related to the reduced-complexity structure $B$ in Fig.~\ref{Fig:HybridStructure}, where a (possibly static) analog precoder is applied to a subset of an antenna array to reduce the training overhead. The beamformed approach may assume the full-complexity structure $A$ in Fig.~\ref{Fig:HybridStructure}, where analog beamformers are used for downlink training signals.
The BS transmits multiple analog precoded pilots in different time or frequency resources. Then, user feedback indicates the preferred analog beam; given this, the user can further measure and feedback the instantaneous effective channel in a legacy LTE manner. These approaches can under some circumstances reduce the overhead in average CSI acquisition, which generally perform well for SU-MIMO but may suffer large performance degradation for MU-MIMO unless the average CSI of all users is fed back. Recently proposed hybrid CSI acquisition schemes in 3GPP combine the above two approaches. First, the BS sends non-precoded pilots to estimate the average CSI at users. Then, based on the analog (non-codebook-based) or digital (codebook-based) feedback of the average CSI from users, the BS determines the analog beamformer and then sends beamformed pilots. These hybrid schemes essentially enable the form of beamforming we discussed above in this section, namely adaptation of the analog beamformer based on long-term statistics, which is then followed by digital beamformer based on instantaneous effective CSI. Increasing the array size further motivates studies to reduce the training and feedback overhead through, e.g., aperiodic training schemes. The JSDM-based structure $C$ in Fig.~\ref{Fig:HybridStructure} that separates a cell into multiple ``virtual sectors'' is one approach to reduce the overhead significantly by simultaneous downlink training and uplink feedback across virtual sectors.

\section{Hybrid Beamforming with selection}
A special class of hybrid systems involves a selection stage that precedes (at the TX) or succeeds (at the RX) the analog processing, called hybrid beamforming with selection hereinafter. 

The up-converted data streams at the TX pass through the analog precoder $\mathbf{F}_{\rm RF}$, as discussed. However, unlike conventional hybrid beamforming, the number of input ports of the analog block is $L \ge N_{\rm RF}^{\rm BS}$ (and typically, $L=N_{\rm BS}$). A selection matrix $\mathbf{S}$, realized by a network of RF switches, feeds the data streams to the best $N^{\rm BS}_{\rm RF}$ out-of-the $L$ ports for transmission. The premise for such a design is that, unlike switches, analog components like phase shifters and amplifiers might not be able to adapt to the quick variation of instantaneous channels over time. Therefore, $\mathbf{F}_{\rm RF}$ is either fixed or designed based on average channel statistics as described in Section~III and $\mathbf{S}$ picks the best ports for each channel realization. 
The switching networks are also advantageous in comparison to full-complexity analog beamforming circuitries in terms of their cost and energy efficiency~\cite{Rial16_Access}. Though we focus on the TX for brevity, a switched analog combiner may also be implemented at the RX.

\subsection{Design of analog precoding/combining block}
\begin{figure}[!htb]
\centering
\includegraphics[width=0.6\textwidth]{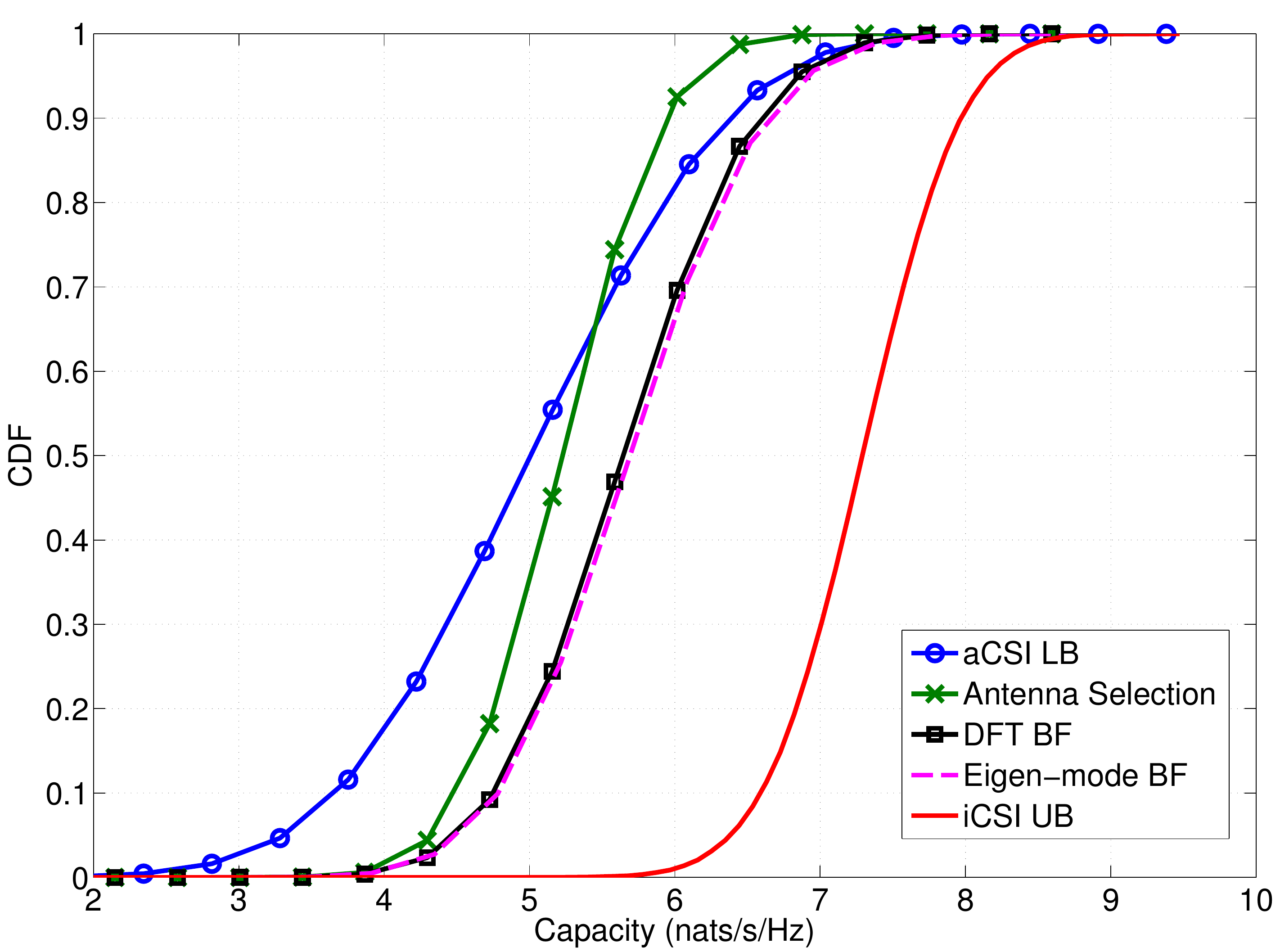}
\caption{Performance of different analog precoders in a hybrid TX with selection. We consider a SU-MIMO system at cm-waves with ideal hardware conditions, where the RX has full complexity with $N_{\rm UE} = N^{\rm UE}_{\rm RF} = 2$ and the TX has a switched hybrid beamforming structure with $N_{\rm BS} = L = 10$, $N^{\rm BS}_{\rm RF} = 2$. The channels are Rayleigh distributed in amplitude, doubly spatially correlated (both at TX and RX) and follow the Kronecker model of spatial correlation ${[\mathbf{R}_{\rm BS}]}_{ij} = {[\mathbf{R}_{\rm UE}]}_{ij} = {0.5}^{|i-j|}$; `aCSI LB' and `iCSI UB' refer to optimal unconstrained precoding with average CSI~\cite{Sudarshan_et_al_2006} and with instantaneous CSI (see Sec.~\ref{sec_2A}), respectively. The RX SNR is $10$dB.}
\label{Fig_switching_performance}
\end{figure}

The simplest ``hybrid beamforming with selection" performs the antenna selection and omits the analog precoding~\cite{Molisch04_MM, Review_love, Magazine_nosratinia}. However, significant beamforming gains can be achieved by introducing analog precoding before the selection to take advantage of the spatial MPCs. Such an architecture performs signal processing in the beam-space. 
A design for $\mathbf{F}_{\rm RF}$ based on the Discrete Fourier Transform (DFT) was proposed in \cite{Molisch_DFTalt} and further analyzed in \cite{Ayach12_SPAWC, Bai11_IT, Choi06_TWC}. Another design can be obtained by eigenmode beamforming based on the TX correlation matrix~\cite{Sudarshan_et_al_2006}, which shows better performance in correlated channels. 
A generalized design for the precoder, where $\mathbf{F}_{\rm RF}$ may be non-unitary with $L \leq N_{\rm BS}$, was proposed in \cite{Ratnam17_ICC}. Impact of channel estimation overhead on precoder design is considered in \cite{Ratnam16_Globecom}. 
To reduce the CSI feedback overhead for FDD systems, $\mathbf{F}_{\rm RF}$ in the conventional hybrid beamforming can be chosen from a set of predetermined codebook of matrices. By regarding the codebook entries as realizations from switch positions, this design can be interpreted as hybrid beamforming with selection. The codebook design is discussed, e.g., in \cite{Hur13_TCOM, Song15_ICC, Singh15_TWC}. The performance of some of these analog precoders is compared in Fig.~\ref{Fig_switching_performance}. 

\subsection{Design of Selection matrix}
Since complexity of searching for the best ports in the analog block is exponentially increasing with $N_{\rm RF}$, many algorithms have been proposed to reduce it. Several such algorithms have been proposed for antenna selection, which are also applicable to other precoder designs (see \cite{Molisch04_MM} and references therein). 
More recently, several greedy algorithms have been proposed to leverage diversity and spatial-multiplexing gains in a multi-user scenario~\cite{Amadori14_PIMRC, Amadori15_TCOM, Wang15_ICC}. 
Restricted selection architectures allow each RF chain to choose from a subset of the analog ports, whereby reducing both search and hardware complexities~\cite{Roh14_CM}. Iterative algorithms with varying search complexities from a linear to sub-exponential order have been proposed \cite{Rahman14_Globecom}.
An alternative technique that does not use the instantaneous CSI is called eigen-diversity beamforming~\cite{Choi08_TCOM, Choi14_CommLetters}. It draws the selection matrix for each channel realization from an optimized probability distribution, thereby leveraging the temporal diversity.

\section{Hybrid Beamforming at Mm-Wave}
Hybrid beamforming architectures and algorithms in the {cm-wave}-band described in the previous sections can in principle be used at mm-wave frequencies. In practice, however, propagation channel and RF hardware aspects are significantly different in those bands, and hence novel hybrid beamforming techniques taking into account the practicalities are needed. 
At mm-wave frequencies,  
the multipath channel experiences higher propagation loss~\cite{Sun14_CM, Nguyen15_VTCSpring, Samimi13_VTCSpring, Rappaport15_TCOM, Maltsev09_JSAC, Zhang14_GLOBECOM, Akdeniz14_JSAC}, which needs to be compensated by gain from antenna arrays at the TX, RX, or both. While such arrays have reasonable physical size thanks to short wavelengths, fully-digital beamforming solutions become infeasible and hybrid beamforming becomes harder due to power- and cost-related RF hardware constraints~\cite{Poon12_ProcIEEE, Choi06_IEEE, Rizvi08_RWS}.  
Moreover, mm-wave channels may be sparser, such that fewer spatial degrees of freedom is available. The sparsity can be exploited for optimizing channel estimation and beam training.

\subsection{Hybrid beamforming methods exploiting channels' sparsity}

Exploiting the channels' sparsity, {the simplest form of hybrid beamforming in SU-MIMO systems focuses array gains to a limited number of multipaths in the RF domain, while multiplexing data streams and allocating powers in baseband. This hybrid architecture is asymptotically optimum in the limit of large antenna arrays \cite{Ayach12_SPAWC}. 
For systems with practical sizes of arrays, which for example have 64 to 256 elements for the BS and under 20 elements for the UEs~\cite{Maltsev15_PIMRC, Cudak14_Globecom, Rebeiz15_IMS}, hybrid beamforming structures are highly desirable. In addition, reduction of the hardware and computational complexity is of great interests. For those purposes, a number of hybrid beamforming methods has been proposed for mm-wave SU-MIMO channels that can be categorized into 1) the use of codebooks, 2) spatially sparse precoding, 3) antenna selection, and 4) beam selection.

\begin{itemize}
\item \textit{Use of codebooks:} While having the same principle as the schemes described in Sec.~II-B, the codebook-based beamforming does not directly estimate the large CSI matrix at the RX, but instead it performs downlink training using pre-defined beams and then only feeds back the selected beam IDs to the transmitter~\cite{Roh14_CM, Singh15_TWC}. 
To further reduce the complexity of beam search and feedback overhead for large antenna systems, a codebook for full-complexity hybrid architecture can be designed to exploit the sparsity of mm-wave channels~\cite{Song15_ICC}. Each codeword is constructed based on the Orthogonal Matching Pursuit (OMP) algorithm to minimize the MSE with the pre-defined ideal beam pattern.

\item \textit{Spatially sparse precoding:} This method finds the approximation of the unconstrained (i.e., fully-digital) beamformer as described in Sec.~II-A; at mm-wave bands with electrically large arrays and a small number of dominant multipaths, the approximation can be made sufficiently close to the optimal precoder using a finite number of antenna elements in the array~\cite{25-ElAyach2014, Alkhateeb14_JSTSP}. The multipath sparsity restricts the feasible analog precoders $\mathbf{F}_{\rm RF}$ to a set of array response vectors, and the baseband precoder optimization can be translated into the sparsity-constrained matrix reconstruction, i.e., the cardinality constraint on the number of RF chains. The near-optimal solution of $\mathbf{F}_{\rm BB}$ can then be found using sparse approximation techniques, e.g., OMP~\cite{Tropp07_TIT}. The SE comparison of this method with unconstrained fully-digital beamforming and analog-only beamsteering with perfect transmit CSI is shown in Fig.~\ref{fig:imperfectRF}. 

\item \textit{Antenna selection:} While the general structure is the same as the one for {cm-wave}s described in Sec. IV, in sparse mm-wave channels, fast and greedy antenna subset selection performs as robust as exhaustive antenna search~\cite{Gharavi04_TSP, MendezRizal15_ITAW}. Hybrid antenna selection can outperform a sparse hybrid combiner with coarsely quantized phase shifters in term of power consumption, when both have the same SE performance~\cite{MendezRizal15_ITAW}. There is still a large gap in SE between the hybrid combiner with switches and fully-digital one with ideal phase shifters.

\item \textit{Beam selection:} Another hybrid beamforming structure is based on continuous aperture phased (CAP)-MIMO transceivers. It uses a lens antenna instead of the phase shifters or switches for RF beamforming, and realizes the beamspace MIMO (B-MIMO) \cite{Sayeed10_Allerton, Brady13_TAP} similarly to the spatial DFT with selection discussed in Sec.~IV-A. 
An electrically-large lens antenna is excited by a feed antenna array beneath the lens. The feed array is called a beam selector since the lens antenna produces high-gain beams that point different angles depending on the feed antenna. The CAP-MIMO can efficiently utilize the low-dimensional high-gain beamspace of the sparse multipath channel by selecting a couple of feed antennas using a limited number of RF chains, like the spatially sparse precoding. 

\end{itemize}

\subsection{Hybrid beamforming in mm-wave MU scenarios}
Hybrid beamforming is also a promising solution for mm-wave MU-MIMO systems. The hybrid structure at the BS can transmit multiplexed data streams to multiple UEs; each UE can be equipped with an antenna~\cite{Bogale16_TWC} or an antenna array with fully-analog beamforming~\cite{07-Liang2014, Han15_CM, Bogale16_TWC, Zheda2016, Alkhateeb2015}. 

Fig.~\ref{Fig:coverage} shows achievable rates of hybrid beamforming in MU multi-cell scenarios~\cite{Alkhateeb2015}. Consider UEs with a single RF chain and many antennas, which distributively select the strongest beam pair to construct analog beamformers. Thanks to the ZF digital precoding at the BS mitigating the inter-user interference, the hybrid structure outperforms significantly the analog beamsteering approach. 

  \begin{figure}[!htb]
    \centering
    \includegraphics[width=0.45\textwidth]{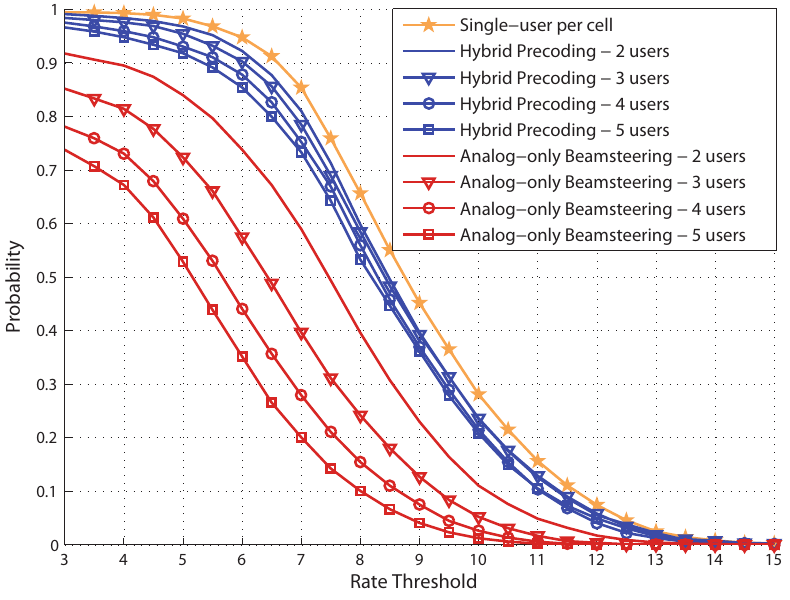}
    \caption{Comparison of achievable rates for hybrid precoding and analog-only beamsteering, from~\cite{Alkhateeb2015}. A single-path model is assumed between the BSs and UEs, and each link is assigned a line-of-sight or non-line-of-sight condition based on a blockage model, i.e., the second reference in~\cite{Alkhateeb2015}. Each UE is associated to the BS with the least path-loss and the BS randomly selects $n = 2, . . . ,5$ associated UEs to be simultaneously served.}
    \label{Fig:coverage}
\end{figure}

The hybrid beamforming based on beam selection and B-MIMO concept can also be extended to the MU-MIMO systems with linear baseband precoders~\cite{Sayeed2013,Amadori14_PIMRC, Amadori15_TCOM}. While its effectiveness (compared to the full-complexity counterparts) has been demonstrated in mm-wave channels, many system and implementation aspects of hybrid beamforming in mm-wave MU-MIMO systems, including multi-user scheduling, and 2D and 3D lens array design, are still open for further research.

\subsection{Impact of transceiver imperfections}
\label{section:imperfection}
The presence of RF transceiver imperfections degrades SE in various ways: e.g., it is harder to accurately generate desired transmit signals when higher beamformer gain is aimed for; non-linear distortion at the RX depends on the instantaneous channel gain and hence the SNR~\cite{Bjornson14_TIT}. Previous works on hybrid beamforming at the cm-wave-band suggest that the hybrid structure achieves the same spectral-efficiency performance as the fully-digital beamforming if the number of RF chains at each end is equal or greater than twice the number of data streams \cite{Zhang05_TSP, 20-Sohrabi2015}. Due to the transceiver imperfections being more pronounced at mm-waves, the SE and SNR of hybrid precoder/combiners no longer scales well with the number of RF chains. Fig. \ref{fig:imperfectRF} compares the SE of spatially sparse hybrid precoding, including RF imperfections, with that from fully-digital beamforming based on SVD. The aggregate impact of the transceiver imperfections is modeled as a Gaussian process~\cite{Studer10_WSA}. The coarsely quantized phase shifters and the transceivers' imperfections significantly degrade the SE. Knowledge of transceiver imperfections at mm-waves is essential for analyzing the scalability of the SE in the large MIMO regime.

\begin{figure}[!htb]
\begin{center}
	    \includegraphics[width=0.55\textwidth]{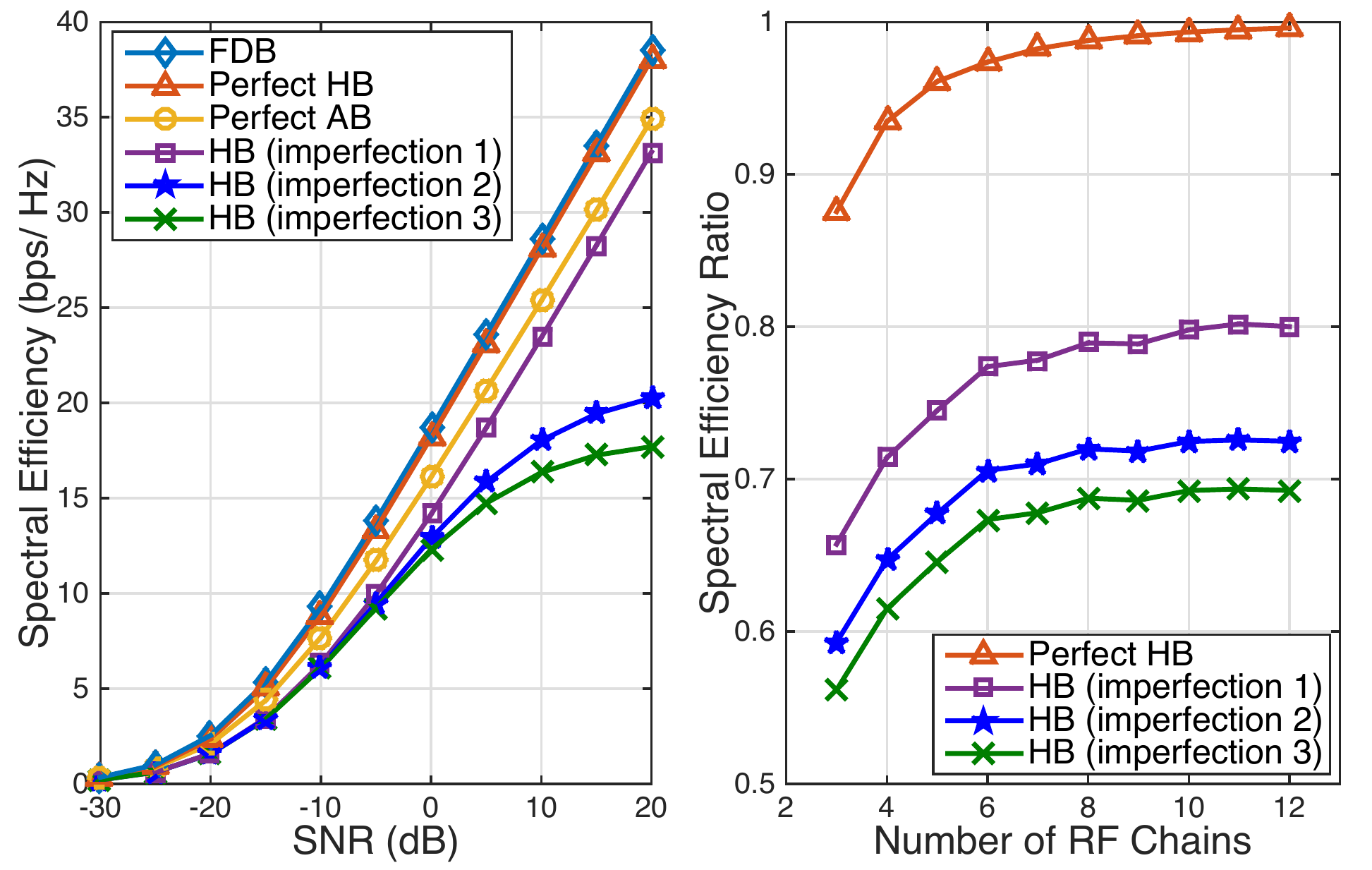}
\caption{SE comparison of Fully-digital Beamforming (FDB), Hybrid Beamforming with perfect RF hardware (Perfect HB), Analog-only Beamsteering with perfect RF hardware (Perfect AB), and HB with three different cases of RF hardware imperfection: case 1 considers quantization error caused by 6-bit phase shifters; cases 2 and 3 additionally consider residual transceiver impairments at BS and at both BS and UE, respectively. The spatially sparse precoding \cite{25-ElAyach2014} is used in the HB. We assume that $N_{\rm BS} = 64$, $N_{\rm UE} = 16$, $N_{\rm S} = 3$, the radio channel has 3 multipath clusters and each has 6 rays, as representative of mm-wave channels. The residual transceiver impairments at TX and RX are characterized by error-vector magnitude of $-20$ dB~\cite{Studer10_WSA}. In the left subfigure, $N^{\rm BS}_{\rm RF} = N^{\rm UE}_{\rm RF} = 6$. In the right subfigure, the SE of the HB with different RF hardware assumptions normalized to the FDB is characterized at $\text{SNR} = 0\ \text{dB}$ and $N^{\rm BS}_{\rm RF} = N^{\rm UE}_{\rm RF}$.}
\label{fig:imperfectRF}
\end{center}
\end{figure}

\subsection{Spectral-energy efficiency tradeoff}
\label{section:se-ee}
Finally, we discuss a relationship between energy efficiency (EE)-SE of hybrid beamforming structures at mm-waves based on~\cite{Han15_CM}. The hybrid structure B in Fig.~\ref{Fig:HybridStructure} was studied, where the BS uses a sub-array with $N_{\rm BS}/N_{\rm RF}^{\rm BS}$ antennas to serve each user individually. Figure~\ref{Fig:EE-SE} shows the EE-SE tradeoff, indicating an optimal number of RF chains achieving the maximal EE for any given SE. 

  \begin{figure}[!htb]
    \centering
    \includegraphics[width=0.45\textwidth]{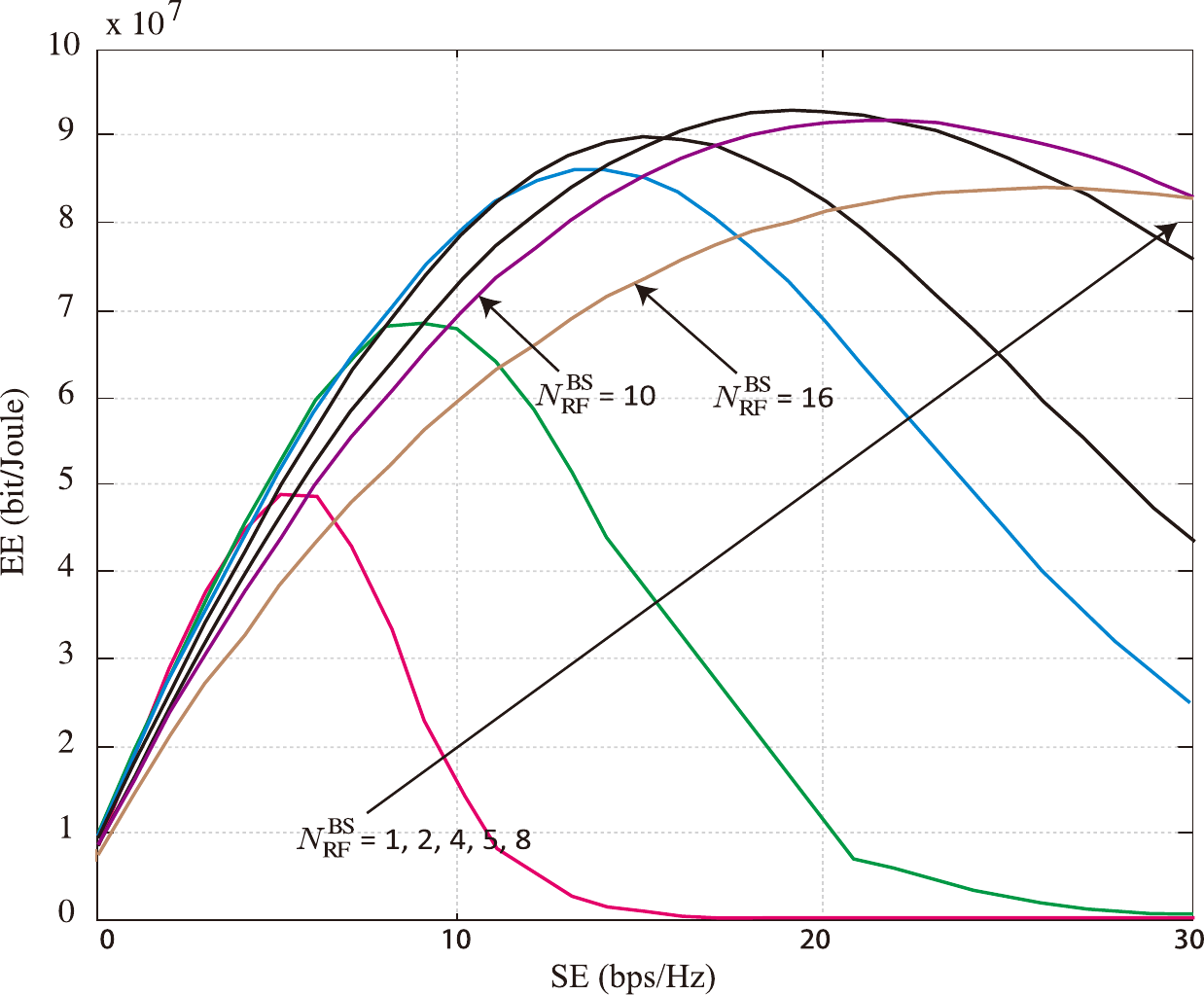}
    \caption{EE-SE relation of mm-wave massive MIMO system~\cite{Han15_CM}. The hybrid transceiver follows Structure B of Fig.~\ref{Fig:HybridStructure}; $N_{\rm BS} = 800$, system bandwidth is 200~MHz, noise power spectral is $10^{-17}$~dBm/Hz, average channel gain is $-100$~dB, the efficiency of power amplifier is 0.375, the static power consumptions for each RF chain and each antenna are both 1~Watt, and the other fixed power consumption is 500~Watt.}
    \label{Fig:EE-SE}
  \end{figure}
  
\section{Conclusion}

Hybrid beamforming techniques were invented more than 10 years ago, but have seen a dramatic uptick in interest in the past 3 years, due to their importance in making massive MIMO systems cost- and energy-efficient. They use a combination of analog and digital beamforming to exploit the fine spatial resolution stemming from a large number of antenna elements, yet keep the number of (expensive and energy-hungry) RF up/downconversion chains within reasonable limits. This paper categorized the hybrid beamforming according to (i) amount of required CSI (instantaneous versus average) for the analog beamformer part; (ii) complexity (full complexity, reduced complexity, and switched), and (iii) carrier frequency range ({cm-wave} versus mm-wave, since both channel characteristics and RF impairments are different for those frequency ranges). It is clear that there is no single structure/algorithm that provides the ``best" tradeoff between complexity and performance in all those categories, but rather that there is a need to adapt them to application and channel characteristics in every design. 

\section*{Acknowledgement}

The financial support of the Academy of Finland and the National Science Foundation through the WiFiUS project ``Device-to-Device Communications at Millimeter-Wave Frequencies" is gratefully acknowledged. 

\bibliographystyle{IEEEtran}
\bibliography{IEEEabrv,ref,references_beamsel,hybrid_beamforming_mmwave,Ref_hybrid1,Ref_hybrid,MassiveMIMO}
\end{document}